\documentclass[conference]{IEEEtran}
\IEEEoverridecommandlockouts
\usepackage{tikz}
\usetikzlibrary{positioning}
\usepackage{cite}
\usepackage{amsfonts}
\usepackage{algorithmic}
\usepackage{textcomp}
\usepackage{xcolor}

\usepackage{url}
\usepackage{cite}
\usepackage{ifthen}
\usepackage[cmex10]{amsmath}
\usepackage{amssymb,bm}
\usepackage{amsthm}

\usepackage{graphics,graphicx,color}
\usepackage{upgreek}
\usepackage[caption=false]{subfig}
\usepackage{pgfplots}
\usepackage{pgfplotstable}
\usepackage[dvipsnames]{xcolor}

\usepackage{balance}

\pgfplotsset{compat=1.18} 

\usepackage{mathtools}
\usepackage{subfig}
\usepackage{microtype}
\usepackage{booktabs}
\usepgfplotslibrary{groupplots}
\usepgfplotslibrary{fillbetween}
\usepackage{tikz}
\usetikzlibrary{plotmarks}
\usepackage{float}
\usepackage{placeins}
\usepackage{physics}

\newcommand{\ind}{\mathbf{1}}
\newcommand{\GF}{\mathrm{GF}}
\newcommand{\eps}{\varepsilon}

\newcommand{\FER}{\mathrm{FER}}

\newcommand{\atanh}{\operatorname{atanh}}

\newcommand{\J}{\mathit{J}}
\newcommand{\epsfromLzeroBPtwo}[1]{%
  \pgfmathparse{3/(2*(exp(#1)+1))}%
  \pgfmathprintnumber[fixed,precision=3]{\pgfmathresult}%
}
\usepackage{newtxtext,newtxmath}
\usepackage{cmap}
\usepackage[utf8]{inputenc} 
\usepackage[T1]{fontenc}
\newcommand{\epsfromLzeroBPfour}[1]{%
  \pgfmathparse{3/(exp(#1)+3)}%
  \pgfmathprintnumber[fixed,precision=3]{\pgfmathresult}%
}
\definecolor{paars}{RGB}{66,20,95}
\definecolor{robijnrood}{RGB}{202,0,93}
\definecolor{geel}{RGB}{255,200,46}
\theoremstyle{definition}

\begin{document}
\title{On LLR Mismatch in Belief Propagation Decoding of Overcomplete QLDPC Codes
\thanks{This research was partly funded by the Dutch Ministry of Defence via the Purple Nectar Quantum Challenge 2025 program (Project TORQUE).}
}
%%%%%
\author{
\IEEEauthorblockN{
Hernan Cordova, 
Alexios Balatsoukas-Stimming, 
Gabriele Liga, 
Yunus Can Gültekin, 
and~Alex Alvarado
}
\IEEEauthorblockA{
Department of Electrical Engineering,
Eindhoven University of Technology, The Netherlands
}

\IEEEauthorblockA{
\texttt{\{h.x.cordova, a.k.balatsoukas.stimming, y.c.g.gultekin, g.liga, a.alvarado\}@tue.nl}
}
}
%%%%
\maketitle
%%%%
\begin{abstract}
Belief propagation (BP) decoding of quantum low density parity check (QLDPC) codes is often implemented using overcomplete stabilizer (OS) representations, where redundant parity checks are introduced to improve finite length performance. Decoder behavior for such representations is governed primarily by finite iteration dynamics rather than asymptotic code properties. These dynamics are known to critically depend on the initialization of the decoder. In this paper, we investigate the impact of mismatched log likelihood ratios (LLRs) used for BP initialization on the performance of QLDPC codes with OS representations. Our results demonstrate that initial LLR mismatch has a strong influence on the frame error rate (FER), particularly in the low noise regime. We also show that the optimal performance is not sharply localized: the FER remains largely insensitive over an extended region of mismatched LLRs. This behavior motivates an interpretation of LLR mismatch as a regularization control parameter rather than a quantity that must be precisely matched to the quantum channel. 
\end{abstract}

%%%%%%%%%%%%%%%%%%%%%%%%%%%%%
%%%%%%%%%%%%%%%%%%%%%%%%%%%%%
\section{Introduction}

Quantum computing promises to transform fields such as drug discovery, materials science, cryptography, and optimization by tackling problems that are intractable for classical computers. However, quantum systems are inherently fragile: qubits are prone to noise, decoherence, and operational imperfections that rapidly corrupt quantum information. Robust quantum error correction (QEC) is therefore indispensable. Without QEC, scaling beyond today's small noisy prototypes is impossible. Continued advances in QEC will be key to transitioning from proof of concept devices to early fault tolerant architectures and, ultimately, to practical large scale quantum computing. One particularly promising option for QEC is quantum low density parity check (QLDPC) codes \cite{Panteleev_2021,evra2020decodablequantumldpccodes,Tillich_2014}. %% add some generic references to QLDPC codes --> DONE

Belief propagation (BP) decoding is widely known for classical LDPC codes and was introduced in the context of QLDPC codes in~\cite{Mackay2004,Hagiwara2007, PoulinChung2008,BP_decodingQLDPC_2024_10619083}.
In its quaternary formulation (BP4), BP is implemented on $\GF(4)$ and operates directly on Pauli symbols. BP4 naturally accounts for degeneracy in quantum error correction
\cite{LeiferPoulin2008,PoulinChung2008}.
Recent decoder implementations have shown that performance can be improved by operating
BP on overcomplete stabilizer (OS) representations, in which redundant parity checks are explicitly
included in the Tanner graph
\cite{smiao_qbp_overcomplete2025,cumitini2024optimalsingleshotdecodingquantum}. As an alternative to BP4, decoding of QLDPC codes can also be performed using BP2, based on binary projections of the stabilizer matrix.

While overcompleteness can reduce frame error rates (FERs) at finite block lengths, it also significantly alters the graph structure on which BP operates.
Redundant OSs increase the density of short cycles, particularly length-$4$ cycles, which strengthen
message correlations and invalidate locally tree like assumptions and traditional asymptotic analysis techniques such as density evolution\cite{RichardsonUrbanke2008,Montanari2005}. 
As a result, decoder behavior on OSs is governed primarily by finite iteration
dynamics rather than asymptotic convergence properties. 

The assumed channel parameter (e.g., the depolarization probability $\eps_{0}$ in a depolarizing channel) does not merely represent a channel parameter but directly determines the magnitude of the initial log likelihood ratio (LLR) injected into the decoder. This initialization, in turn, governs how rapidly messages saturate and how strongly information is reinforced through short feedback loops, particularly through length-$4$ cycles that are abundant by design in OS graphs. As a result, the assumed channel parameter acts as an implicit regularization or stability control parameter for early BP dynamics rather than as an asymptotically optimal choice. 

For BP4 decoders operating on OS graphs, it has been empirically observed in \cite[Fig. 3]{smiao_qbp_overcomplete2025} that moderate channel parameter mismatch can reduce FERs in the moderate (to high) noise regime. This mismatch corresponds to the case when the assumed (initial) channel parameter differs from the true channel parameter. While the observations of mismatch reducing FER for BP4 are new in the quantum setup, these were originally suggested for classical LDPC codes over binary symmetric channels in\cite{Hagiwara2012}. 
Current results in the literature do not provide a clear answer regarding the operational role of channel mismatch for QLDPC codes, nor do they clarify whether such sensitivity is specific to quaternary BP (BP4) or reflects a more general property of BP on OS graphs.
 
In this paper, we present a detailed study of the aforementioned LLR mismatch.  We focus on both BP4 and BP2 decoding.  
Our main contribution is to show a strong dependency of the performance of both decoders on LLR mismatch. However, our results also show that the optimum mismatch is not sharply localized. Instead, the apparent optimum corresponds to an operational regime of comparable mismatch, rather than a single finely tuned value. These results motivate us to interpret the LLR calculation as a finite iteration regularization parameter rather than as a process that must be perfectly matched to the quantum channel. Our numerical results show that in the low noise regime, mismatched LLRs provide gains of nearly two orders of magnitude with respect to the matched counterparts, both for BP2 and BP4. To the best of our knowledge, this paper shows for the first time that LLR mismatch offers large FER gains, and that these gains are observable for both BP2 and BP4.

\textbf{\textit{Notation}}: 
Scalars are lowercase (e.g., $\eps$), vectors bold (e.g., $\boldsymbol{E}$), and matrices uppercase (e.g., $H$).
Finite fields are denoted by $\GF(2)$ and $\GF(4)$.
Calligraphic letters denote index sets (e.g., $\mathcal{G}$).
The trace inner product is $\langle\cdot,\cdot\rangle_{\mathrm{tr}}:\GF(4)^n\times\GF(4)^n\to\GF(2)$.
Addition in $\GF(4)^n$ is written as $\oplus$ and corresponds to the component wise Pauli multiplication modulo global phase. $\FER(\eps;L_0)$ denotes the frame error rate.
%%%%%%%%%%%%%%%%%%%%%%%%%%%%%%%%%%%%%%%%%%%%%%%%%%%%%%%%%%%%%%%
\section{System Model and BP Decoders}
\label{sec:model_decoders}

Fig.~\ref{fig:qecblock} shows an overview of a QEC-protected quantum system under consideration. First, $k$ logical qubits are encoded into $n$ physical qubits denoted by $\ket{\psi}$. The encoded qubits are affected by a quantum channel defined by a depolarizing probabily $\eps$. This channel introduces noise $\boldsymbol{E}$, resulting in a noisy output $\boldsymbol{E} \ket{\psi}$, where $\boldsymbol{E}$ acts component wise on the physical qubits. The BP decoder uses the $m$-bit syndrome vector obtained through stabilizer measurements on the noisy qubits $\boldsymbol{E}\ket{\psi}$ as well as an LLR, $L_{\mathrm{ch}}(\eps_{0})$, which is a function of the assumed channel parameter $\eps _0$, to produce an estimate $\hat{\boldsymbol{E}}$ of the error vector $\boldsymbol{E}$. In the following subsections, we describe the aforementioned steps in more detail.

\subsection{Quantum LDPC Codes}
We consider a QLDPC code encoding $k$ logical qubits into $n$ physical qubits and specified by a parity check matrix $H\in\GF(4)^{m\times n}$ satisfying the symplectic orthogonality
conditions required for valid stabilizer generators
\cite{Gottesmanqecfault2010,Tillich_2014}.
Encoding is performed with CNOT gates between ancillary qubits and $k$ logical qubits, as in~\cite{Gottesmanqecfault2010}.

%%%%%%%%%%%%%%%%%%%%%%%%%%%%%
\subsection{Quantum Channel and Stabilizers}
%%%%%%%%%%%%%%%%%%%%%%%%%%%%%
We assume a depolarizing channel, where each qubit $j$ is affected  independently and identically. The error $E_j$ on qubit $j$ is modeled as a Pauli operator $E_j\in\{I,X,Y,Z\}$. For decoding purposes, we represent these Pauli operators by their corresponding elements in $\GF(4)$ under the  standard stabilizer isomorphism described in~\cite{Gottesman1997}. Throughout, the symbols $\{I,X,Y,Z\}$ denote both the physical Pauli matrices and their associated $\GF(4)$ representatives, with the identification understood from the context.
Under the depolarizing channel, the $n$-qubit error pattern is
$\boldsymbol{E}=(E_1,\dots,E_n)\in\GF(4)^n$ and the errors are distributed as 
%%%%%%%%%%%%%%%%%%%%%%%%%%%%%
\begin{equation}
\Pr\{E=e\} = 
\begin{cases}
1-\eps, & e = I\\ 
\eps/3, & e =X\\
\eps/3, & e =Y\\
\eps/3, & e =Z
\end{cases}\,,
\label{eq:depol_sys}
\end{equation}
%%%%%%%%%%%%%%%%%%%%%%%%%%%%%
where $\eps$ is the depolarization probability and where we used $\Pr\{E_j=e\}=\Pr\{E=e\}$ because the channel is memoryless.

%%%%%%%%%%%%%%%%%%%%%%%%%%%%%
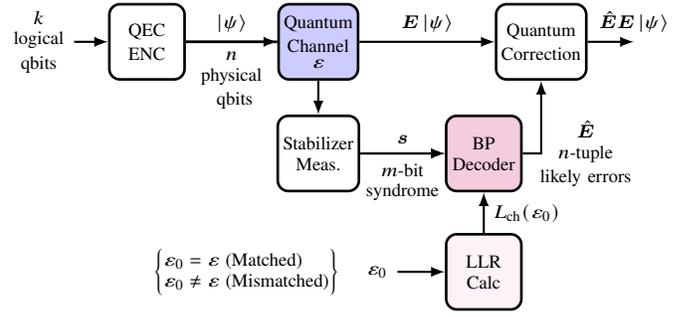
\begin{figure}[!t]
    \centering
    \begin{tikzpicture}[line width=1pt, 
font=\scriptsize,
block/.style={rectangle, rounded corners,draw,inner sep=2pt,
minimum width=10mm, minimum height=10mm,align=center},
linestyle/.style = {-latex,thick}
]

\def\dist{0.90}

\node[align=center] at (0,\dist) (start)
    {\shortstack{$k$\\ logical \\  qbits}};
%%%%%%%%%%%%%%%
% QEC ENC
\node[block, right=0.5*\dist of start] (ENC)
    {\shortstack{QEC \\ ENC}};
\draw[linestyle] (start) -- (ENC.west);
%%%%%%%%%%%%%%%
% Quantum Channel 
\node[block, right=1.35*\dist of ENC, fill=blue!20] (channel)
    {\shortstack{Quantum \\ Channel \\ $\eps$}};

\draw[linestyle] (ENC) -- (channel.west)
    node[pos=0.5,below] {\shortstack{$n$  \\  physical \\\ qbits}};
\draw[-] (ENC) -- (channel.west)
    node[pos=0.5,above] {$\ket{\psi}$};
 %%%%%%%%%%%%%%%   
%%%% Quantum Correction
\node[block, right=2.0*\dist of channel] (correct)
    {\shortstack{Quantum \\ Correction}};
\draw[linestyle] (channel) -- (correct.west)
    node[pos=0.5,above] {$\boldsymbol{E}\ket{\psi}$};
%%%%%%%%%%%%%%%
%%%%% Stabilizer Meas.
%\node[block, below right=0.5 and 0.8 of channel] (stabile)
\node[block, below=0.5*\dist of channel] (stabile)
    {\shortstack{Stabilizer \\ Meas.}};

%\draw[linestyle] ($(channel.east)+(0.2,0)$) |- (stabile.west);
\draw[linestyle] (channel.south) -- (stabile.north);
%%%%%%%%%%%%%%%
%%%% BP DECODER
\node[block, right=1.25*\dist of stabile, fill=robijnrood!20] (DEC)
    {\shortstack{BP \\ Decoder}};

\draw[linestyle] (stabile) -- (DEC)
    node[pos=0.5,below] {\shortstack{$m$-bit \\ syndrome}};
\draw[linestyle] (stabile) -- (DEC)
    node[pos=0.5,above] {$\boldsymbol{s}$};
    
\draw[linestyle] (DEC.east) -| (correct.south)
    node[pos=0.2,right,align=center]
    {\shortstack{$\hat{\boldsymbol{E}}$ \\ $n$-tuple \\ likely errors}};
%%%%%%%%%%%%%%%%
% LLR Comp
\node[block, below=0.6*\dist of DEC, fill=robijnrood!05] (LLR)
    {\shortstack{LLR \\ Calc}};
% epsilon0 input
\node[left=0.7*\dist of LLR] (eps0)
    {$\eps_{0}$};
\draw[linestyle] (eps0) -- (LLR.west);
% LLR output upward to BP decoder
\draw[linestyle] (LLR.north) -- (DEC.south)
    node[pos=0.5,right] {$L_{\mathrm{ch}}(\eps_0)$};
%%%%%%%%%%%%%%%
% Match / Mismatch Text block (LEFT of eps0)
\node[left=0.01*\dist of eps0, align=left] (matchinfo)
{
\scriptsize
$\left\{
\begin{array}{@{}l@{}}
\eps_{0} =\eps \; \text{(Matched)} \\
\eps_{0} \neq \eps \;  \text{(Mismatched)}
\end{array}
\right\}$
};
%\draw[linestyle] (matchinfo.east) -- (eps0.west);
%%%%%%%%%%%%
\node at ($(correct.east)+(0.75,0)$) (end) {};
\draw[linestyle] (correct) -- (end)
    node[pos=1,above,align=center]
    {$\hat{\boldsymbol{E}}\boldsymbol{E}\ket{\psi}$};

\end{tikzpicture}
    \caption{\footnotesize QEC decoding block diagram. The same OS GB$(126,28,126)$ construction ($k=28$, $n=m=126$) is used for both BP4 and its BP2 equivalent. The true channel depolarizing probability is $\eps$. The LLR computation block uses $\eps_{0}$.}
    \label{fig:qecblock}
\end{figure}
%%%%%%%%%%%%%%%%%%%%%%%%%%%%%
The syndrome vector $\boldsymbol{s}\in \GF(2)^{m}$, given an error realization $\boldsymbol{E}\in \GF(4)^{n}$, is defined component wise as:
\begin{equation}
s_i=
\langle H_i,\boldsymbol{E}\rangle_{\mathrm{tr}},
\qquad i=1,\dots,m,
\label{eq:syndrome_trace}
\end{equation}
where $H_i$ denotes the $i$-th row of $H$ and $\langle\cdot,\cdot\rangle_{\mathrm{tr}}$ is the trace inner product.
Decoding is performed at the coset level, meaning that two error patterns $\boldsymbol{E}$ and $\boldsymbol{E}'$ that differ by a stabilizer element produce identical syndromes and
act identically on the codespace~\cite{Gottesman1997}. Therefore, the decoder seeks an estimate within the correct stabilizer coset rather than the exact physical error.
Let $\hat{\boldsymbol{E}}$ denote the decoder estimate of the error (see the output of the BP decoder block in Fig.~\ref{fig:qecblock}). The residual error is defined as $\tilde{\boldsymbol{E}} = \hat{\boldsymbol{E}} \oplus \boldsymbol{E}$, where $\oplus$ denotes addition in $\GF(4)^{n}$ which corresponds to the component wise Pauli multiplication modulo global phase as explained in~\cite{Gottesman1997}.
A decoding attempt is declared successful if the residual error 
$\tilde{\boldsymbol{E}} \in \mathcal{S}$, 
i.e., if $\tilde{\boldsymbol{E}}$ commutes with all stabilizer generators. $\mathcal{S}$ here denotes the stabilizer group. In practice, this condition is verified using the stabilizer generator matrix $G$, which contains a set of independent generators of the stabilizer group $\mathcal{S}$ that are used to verify the membership of the residual error $\tilde{\boldsymbol{E}}$ in $\mathcal{S}$, as explained in~\cite{cumitini2024optimalsingleshotdecodingquantum, Gottesman1997}.
%%%%%%%%%%%%%%%%%%%%%%%%%%%%%
\subsection{BP Decoders}
We focus on OS representations, in
which additional redundancy is included so that $m>n-k$. 
Variable nodes (qubits) are indexed by $j\in\{1,\dots,n\}$ and check nodes (stabilizers)
by $i\in\{1,\dots,m\}$.
Let ${\mathcal{N}(j)}$ denote the set of checks adjacent to variable node $j$ and
${\mathcal{M}(i)}$ the set of variables adjacent to check node $i$.

BP2 decoding is performed on the binary projections of the stabilizer matrix. 
Defining $(H_Z)_{i,j}\!=\!\ind\{H_{i,j}\in\{X,Y\}\}$ and
$(H_X)_{i,j}\!=\!\ind\{H_{i,j}\in\{Z,Y\}\}$ leads to the binary parity relations $
\boldsymbol{s_Z} = H_Z\boldsymbol{e}_X \pmod{2}$ and $\boldsymbol{s_X} = H_X\boldsymbol{e}_Z \pmod{2}$, respectively, where $\boldsymbol{e}_X$ and $\boldsymbol{e}_Z$ denote the binary vectors indicating the presence of $X$ and $Z$ components in $\boldsymbol{E}$, respectively.
BP2 decoding is applied separately to these two binary systems using a maximum of $\ell^{\max}$ iterations and a 
syndrome-based stopping rule. Since physical noise acts on qubits, only variable nodes are assigned initial LLR values, while check nodes represent deterministic parity constraints and are initialized with LLRs set to zero\cite{smiao_qbp_overcomplete2025}. Further details can be found in\cite{PoulinChung2008, LeiferPoulin2008,BP_decodingQLDPC_2024_10619083}.

For BP4, we consider the same characterization as in\cite{smiao_qbp_overcomplete2025}, which performs scalar message passing instead of (quaternary) vector message passing.
%%%%%%%%%%%%%%%%%%%%%%%%%%%%%
Let $L^{(\ell)}_{j\to i}$ and $L^{(\ell)}_{i\to j}$ denote the variable to check (V2C) and
check to variable (C2V) LLR messages at iteration $\ell$, respectively.
The C2V update at check node $i$ is:
%%%%%%%%%%%%%%%%%%%%%%%%%%%%%
\begin{equation}
L^{(\ell)}_{i\to j}
=
(-1)^{s_i}\,2\,\atanh\!\Bigg(
\prod_{k\in\mathcal{M}(i)\setminus\{j\}}
\tanh\!\left(\frac{L^{(\ell)}_{k\to i}}{2}\right)
\Bigg),
\label{eq:bp4_check_update}
\end{equation}
%%%%%%%%%%%%%%%%%%%%%%%%%%%%%
where $s_i$ is defined as in~\eqref{eq:syndrome_trace} and the check node sign factor $(-1)^{s_i}$ in~\eqref{eq:bp4_check_update} accounts for the measured syndrome value, so that each stabilizer node $i$ reflects whether the cumulative Pauli error in its
neighborhood commutes ($s_i=0$) or anticommutes ($s_i=1$) with the corresponding
stabilizer generator.

The V2C update at each variable node $j$ at iteration $\ell$ is obtained by consolidating all incoming C2V messages, except the destination check, and by adding the initial LLR contribution $L_{\mathrm{ch}}^{\mathrm{BP4}}(\eps_{0})$, as:
\begin{equation}
   L^{(\ell)}_{j\to i}= L_{\mathrm{ch}}^{\mathrm{BP4}}(\eps_{0})+
\sum_{i'\in\mathcal{N}(j)\setminus\{i\}}
L^{(\ell)}_{i'\to j},
\label{eq:bp4_VN_aggregation}
\end{equation}
Following the scalarization principle introduced in~\cite{Lai_2021}
and applied in~\cite{smiao_qbp_overcomplete2025}, 
three a posteriori LLRs  are computed 
at each variable node $j$, corresponding to the $X$, $Y$, and $Z$ 
error (Pauli) hypotheses (relative to the identity hypothesis). These values are obtained from the scalar edge messages without vector message passing. As such, this scalarization technique~\cite{Lai_2021} preserves the relative ordering of Pauli symbol probabilities while reducing computational complexity compared to full vector message BP.
A hard decision selects the Pauli symbol associated with the largest 
a posteriori LLR. If all three LLR values are positive, the identity operator is selected.
The decoder process terminates when the syndrome constraint is satisfied or when the maximum number of iterations $\ell^{\max}$ is reached.

%%%%%%%%%%%%%%%%%%%%%%%%%%%%%
\subsection{LLR Mismatch}
%%%%%%%%%%%%%%%%%%%%%%%%%%%%%
As shown in \eqref{eq:depol_sys}, the quantum (physical) channel is characterized by the depolarization probability $\eps$. 
The BP decoder is initialized at each variable node $j$ with LLRs computed assuming a depolarization probability $\eps_{0}$.  
\emph{Matched} operation corresponds to $\eps_{0}=\eps$, whereas \emph{mismatched} operation refers to the case
$\eps_{0}\neq\eps$.

BP2 decoding runs on Tanner graphs constructed based on binary matrices $H_X$ and $H_Z$.
The messages exchanged are scalar, quantifying the beliefs about errors being $Z$ or $X$, respectively, since $X$ and $Z$ anticommute.
Thus, the initial LLR in BP2 is defined as
\begin{equation}
\begin{aligned}
L_{\mathrm{ch}}^{\mathrm{BP2}}(\eps_{0})
&\triangleq
\underbrace{\ln\!\left(
\frac{\Pr\{I\in\{I, X\}\mid\eps_{0}\}}
{\Pr\{I\in\{Y, Z\}\mid\eps_{0}\}}
\right)}_{\mbox{for} \: H_X}\\
&=
\underbrace{\ln\!\left(
\frac{\Pr\{I\in\{I, Z\}\mid\eps_{0}\}}
{\Pr\{I\in\{X, Y\}\mid\eps_{0}\}}
\right)}_{\mbox{for} \: H_Z}= \ln\!\left(
\frac{1-2\eps_{0}/3}{2\eps_{0}/3}
\right)
\end{aligned}
\label{eq:bp2_e0}
\end{equation}
which follows directly from the error model~\eqref{eq:depol_sys}.

BP4 decoding runs on a Tanner graph constructed based on the quaternary matrix $H$.
The messages exchanged are three-dimensional vectors, quantifying the beliefs about errors being $X$, $Y$, or $Z$.
Thus, the initial LLR vectors in BP4 are defined with identical elements given by
\begin{equation}
L_{\mathrm{ch}}^{\mathrm{BP4}}(\eps_{0})
\triangleq
\ln\!\left(
\frac{\Pr\{E=I\mid\eps_{0}\}}
{\Pr\{E=\xi\mid\eps_{0}\}}
\right)
=
\ln\!\left(
\frac{1-\eps_{0}}{\eps_{0}/3}
\right),
\label{eq:bp4_e0}
\end{equation}
which follows directly from~\eqref{eq:depol_sys} where $\xi \in \{X,Y,Z\}$. 
%i.e., as the log likelihood ratio
%between the identity symbol and any non identity Pauli symbol. 
% The expression in th r.h.s. of \eqref{eq:bp4_e0} also follows directly from~\eqref{eq:depol_sys}.
%%%%%%%%%%%%%%
% For both cases (BP2 and BP4), the LLRs in \eqref{eq:bp2_e0} and  \eqref{eq:depol_sys} determine the initial value of all V2C LLR messages before iterative updates begin.

%%%%%%%%%%%%%%%%%%%%%%%%%%%%%
\section{LLR Mismatch Analysis}
\label{sec:prior_mismatch}
%%%%%%%%%%%%%%%%%%%%%%%%%%%%%

This section formalizes the dependency of finite iteration decoding
performance on the assumed channel parameter $\eps_0$ and introduces a scalar
objective metric to quantify sensitivity to this mismatch. Similar FER observations were reported in \cite{smiao_qbp_overcomplete2025}, but only for BP4 and at relatively high FER values.
%%%%%%%%%%%
%%%%%%%%%%%%%%%%%%%%%%%%%%%%%%%%%%%%%%%%%%%%%%
\begin{figure}[t]
\centering
\begin{tikzpicture}
\definecolor{set1}{RGB}{161,33,33}
\definecolor{set2}{RGB}{69,99,168}
\definecolor{set3}{RGB}{27,158,119}
\definecolor{set4}{RGB}{152,78,163}
\definecolor{set5}{RGB}{255,127,0}
\definecolor{set6}{RGB}{166,118,29}
\definecolor{set7}{RGB}{230,171,2}
\definecolor{set8}{RGB}{231,41,138}
\def\lw{0.75pt}
\begin{axis}[
    yticklabel style={
    /pgf/number format/fixed,
    /pgf/number format/precision=3
    },
    every axis/.append style={font=\footnotesize},
    width=\columnwidth,
    height=0.70\columnwidth,
    xmode=log,
    ymode=log,
    xmin=1e-3, xmax=1e-1,
    ymin=1e-9, ymax=1,
    ylabel near ticks,
    xlabel near ticks,
    xlabel={\scriptsize $\eps$},
    ylabel={\scriptsize FER},
    title={\scriptsize GB code $n=126$, $k=28$, $m=126$, BP2},
    title style={font=\scriptsize},
    grid=major,
    grid style={dashed,lightgray!75},
  %  major grid style={gray!25},
  %  minor grid style={gray!12},
    tick label style={font=\scriptsize},
    legend style={
        font=\scriptsize,
        draw,
        fill=white,
        at={(0.97,0.03)},
        anchor=south east
    },
    legend cell align=left,
]
%%%%%%
\addplot [color=red!50!white, dashed, line width=\lw, mark=o, mark options={solid,scale=0.8, fill=white, line width=1pt}] file[]{figures/tikz_export/FER_BP2/GB_n126_k28_m126_BP2_iter4_match.txt};
\addlegendentry{$\ell^{\max}=4$, match}
%%%%%
\addplot [color=red!50!white, solid, line width=\lw, mark=square*, mark options={solid,scale=0.7, fill=white, line width=1pt}] file[]
{figures/tikz_export/FER_BP2/GB_n126_k28_m126_BP2_iter4_mismatch_ep0_0p10.txt};
\addlegendentry{$\ell^{\max}=4$, mismatch}

\addplot [blue, dashed, line width=\lw, mark=+, mark options={solid,scale=0.8, fill=white, line width=1pt}] file[]
{figures/tikz_export/FER_BP2/GB_n126_k28_m126_BP2_iter8_match.txt};
\addlegendentry{$\ell^{\max}=8$, match}

\addplot [blue,solid, line width=\lw, mark=triangle*, mark options={solid,scale=0.7, fill=white, line width=1pt}] file[]
{figures/tikz_export/FER_BP2/GB_n126_k28_m126_BP2_iter8_mismatch_ep0_0p10.txt};
\addlegendentry{$\ell^{\max}=8$, mismatch}

\end{axis}
\end{tikzpicture}

\caption{\footnotesize
FER for BP2 on the OS GB$(126,28,126)$ code. LLR mismatch (with $\eps_{0}=0.10$) is shown to improve finite iteration FER performance.}
\label{fig:bp2_fer_match_mismatch}
\end{figure}
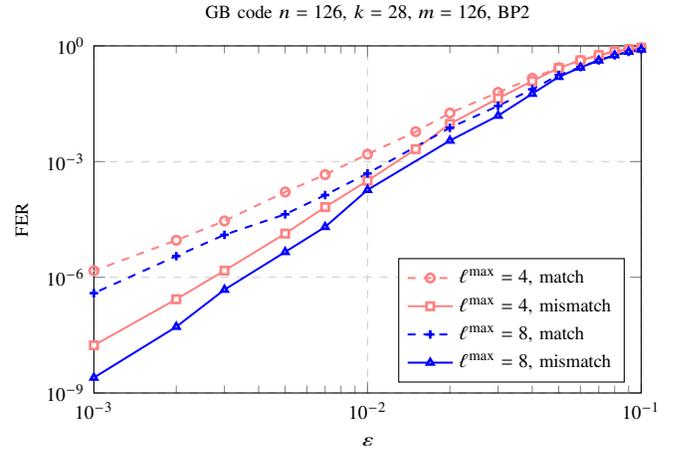
%%%%%%%%%%%%%%%%%%%%%%%%%%%%%%%%%%%%%%%%%%%%%%
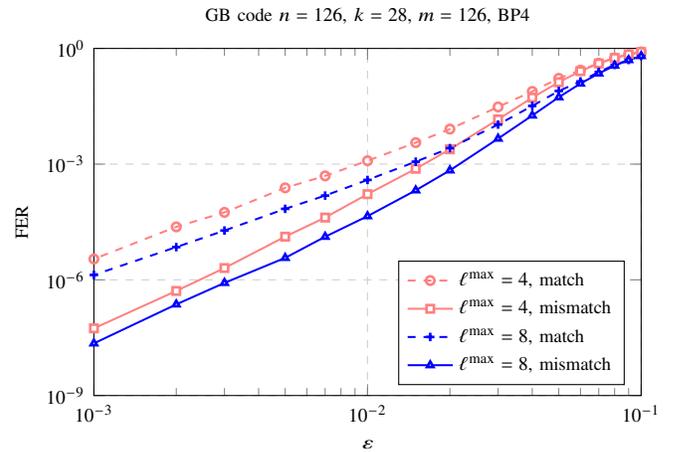
\begin{figure}[t]
\centering
\begin{tikzpicture}
\definecolor{set1}{RGB}{161,33,33}
\definecolor{set2}{RGB}{69,99,168}
\definecolor{set3}{RGB}{27,158,119}
\definecolor{set4}{RGB}{152,78,163}
\definecolor{set5}{RGB}{255,127,0}
\definecolor{set6}{RGB}{166,118,29}
\definecolor{set7}{RGB}{230,171,2}
\definecolor{set8}{RGB}{231,41,138}
\def\lw{0.75pt}
\begin{axis}[
    yticklabel style={
    /pgf/number format/fixed,
    /pgf/number format/precision=3
    },
    every axis/.append style={font=\footnotesize},
    width=\columnwidth,
    height=0.70\columnwidth,
    xmode=log,
    ymode=log,
    xmin=1e-3, xmax=1e-1,
    ymin=1e-9, ymax=1,
    ylabel near ticks,
    xlabel near ticks,
    xlabel={\scriptsize $\eps$},
    ylabel={\scriptsize FER},
    title={\scriptsize GB code $n=126$, $k=28$, $m=126$, BP4},
    title style={font=\scriptsize},
    grid=major,
    grid style={dashed,lightgray!75},
  %  major grid style={gray!25},
  %  minor grid style={gray!12},
    tick label style={font=\scriptsize},
    legend style={
        font=\scriptsize,
        draw,
        fill=white,
        at={(0.97,0.03)},
        anchor=south east
    },
    legend cell align=left,
]
%%%%%%
\addplot [color=red!50!white, dashed, line width=\lw, mark=o, mark options={solid,scale=0.8, fill=white, line width=1pt}] file[]{figures/tikz_export/FER_BP4/GB_n126_k28_m126_BP4_iter4_match.txt};
\addlegendentry{$\ell^{\max}=4$, match}
%%%%%
\addplot [color=red!50!white, solid, line width=\lw, mark=square*, mark options={solid,scale=0.7, fill=white, line width=1pt}] file[]
{figures/tikz_export/FER_BP4/GB_n126_k28_m126_BP4_iter4_mismatch_ep0_0p10.txt};
\addlegendentry{$\ell^{\max}=4$, mismatch}

\addplot [blue, dashed, line width=\lw, mark=+, mark options={solid,scale=0.8, fill=white, line width=1pt}] file[]
{figures/tikz_export/FER_BP4/GB_n126_k28_m126_BP4_iter8_match.txt};
\addlegendentry{$\ell^{\max}=8$, match}

\addplot [blue,solid, line width=\lw, mark=triangle*, mark options={solid,scale=0.7, fill=white, line width=1pt}] file[]
{figures/tikz_export/FER_BP4/GB_n126_k28_m126_BP4_iter8_mismatch_ep0_0p10.txt};
\addlegendentry{$\ell^{\max}=8$, mismatch}

\end{axis}
\end{tikzpicture}

\caption{\footnotesize
FER for BP4 on the OS GB$(126,28,126)$ code. LLR mismatch (with $\eps_{0}=0.10$) is shown to improve finite iteration FER performance.}
\label{fig:bp4_fer_match_mismatch}
\end{figure}
\subsection{LLR Mismatch Impact on BP Decoders}

We consider the OS Generalized Bicycle (GB) $(126,28,126)$ code with $\ell^{\max}=4$ and $\ell^{\max}=8$ decoding iterations. For each physical error rate $\eps$, we generate i.i.d. Pauli errors $E\in\GF(4)^n$ according to \eqref{eq:depol_sys}, compute the syndromes, and decode using either BP2 or BP4. These two decoders used mismatched LLRs with $\eps_{0}=0.10$ so that the presented results are directly  comparable to those reported in~\cite{smiao_qbp_overcomplete2025}.

A matched versus mismatched LLR comparison for BP2 is shown in Fig.~\ref{fig:bp2_fer_match_mismatch}. The results in this figure show that LLR mismatch (with $\eps_{0}=0.10$) has a positive impact on the FER performance of BP2, especially for low FER values. The gains offered by the LLR mismatch are up to two orders of magnitude for a depolarizing channel with $\eps=10^{-3}$. Such gains are visible for both $4$ and $8$ iterations.

Fig.~\ref{fig:bp4_fer_match_mismatch} illustrates the FER performance for BP4, where the decoder is also initialized with a mismatched LLR $\eps_{0}=0.10$. The results in this figure show that the FER relative to the matched operation $\eps_{0}=\eps$ under a fixed number of iterations is again improved. Similar to the BP2 case in Fig.~\ref{fig:bp2_fer_match_mismatch}, the gains are up to two orders of magnitude in the low FER regime. As $\ell^{\max}$ increases from $4$ to $8$, this effect starts to diminish, indicating that LLR mismatch primarily influences transient message evolution rather than ultimate decoding capability.

The results in Figs.~\ref{fig:bp2_fer_match_mismatch}–\ref{fig:bp4_fer_match_mismatch}
do not indicate that LLR mismatch compensates for insufficient iterations,
nor that it improves the decoder’s performance in the limit of large
iteration counts. Instead, they show that under fixed graph structure and iteration budget, finite iteration FER is sensitive to the initialization. % scale and impacts the transient evolution of messages.
This motivates a systematic quantification of initial LLR sensitivity. %over a selected operating region.
The next section proposes a metric to analyze this behavior. Numerical results for this metric are presented in Sec.~\ref{sec:results}.
%%%%%%%%%%%%%%%%%%%%%%%%%%%%

\subsection{Aggregated Objective (AO) Function}

For a fixed physical depolarizing probability $\eps$ and a mismatched LLR, which from now on we denote by $L_0$, the finite iteration frame error rate is defined as
\begin{equation}
\FER(\eps;L_0)
\triangleq
\Pr\{\text{decoder fails}\,\big|\,\eps,L_0\}.
\label{eq:FER_def_clean}
\end{equation}
In practice, $\FER(\eps;L_0)$ is estimated via Monte Carlo as:
\begin{equation}
\FER(\eps;L_0)
\approx
\frac{1}{N(\eps)}
\sum_{r=1}^{N(\eps)}
\ind\!\left(\text{failure at trial } r\right),
\label{eq:FER_hat}
\end{equation}
where $N(\eps)$ is the number of simulated frames.
For a fixed QLDPC code and decoder configuration,
$\FER(\eps;L_0)$ fully characterizes the empirical finite iteration behavior of the decoder as a function of the mismatch LLR $L_0$.

%%%%%%%%%%%%%%%%%%%%%%%%%%%%

To quantify the effect of the mismatched LLR $L_{0}$,
we aggregate performance over a selected (discretized) operating region $\mathcal{G}$ of channel depolarizing probabilities $\eps$, given by
$\mathcal{G} = \{\eps_1,\ldots,\eps_T\}$. We define the Aggregated Objective (AO) as a weighted
geometric mean of the empirical FER as
\begin{equation}
\tilde{\J}(L_0)
=
\prod_{t=1}^{T}
\Big(
\max\{\FER(\eps_t;L_0),x_{\min}\}
\Big)^{w_t},
\label{eq:PAO_geometric}
\end{equation}
where $\{w_t\}_{t=1}^T$ are nonnegative weights satisfying
$\sum_{t=1}^T w_t = 1$, and $x_{\min}$ is a conservative floor
applied when zero frame errors are observed. $\FER$ is computed as in~\eqref{eq:FER_hat}. When $\FER(\eps_t;L_0)=0$, we set
\begin{equation}
x_{\min}
=
1-(0.05)^{1/N(\eps_t)},
\label{eq:CP_floor}
\end{equation}
which corresponds to the one-sided $95\%$ Clopper–Pearson
upper confidence bound\cite{clopper1934confidence} for $N(\eps_t)$ independent trials. This prevents optimistic bias and ensures numerical stability of the AO.
For numerical stability and visualization, we work in the log domain, converting \eqref{eq:PAO_geometric} into
\begin{equation}
\J(L_0)=
\sum_{t=1}^{T}
w_t\,\phi\!\left(\FER(\eps_t;L_0)\right),
\label{eq:PAO_log}
\end{equation}
where $\phi(x) = \log_{10}\!\big(\max\{x,x_{\min}\}\big)$.
Unless otherwise specified, we use uniform weights over the selected
(discretized) set $\mathcal{G}$, 
i.e., $w_t = 1/|\mathcal{G}|=1/T$ for
$\eps_t\in\mathcal{G}$ and $w_t=0$ otherwise. Nonuniform weights can be implemented for example to explore a particular noise regime brought by advances in specific quantum technologies.

Since each assumed mismatched LLR $L_{0}$ produces an entire FER curve over the physical selected noise region $\mathcal{G}$, we compact its overall performance by computing the weighted discrete expectation of the (log) FER in \eqref{eq:PAO_log}. The resulting scalar objective $\J(L_0)$ enables systematic comparison of initial LLRs values and identification of stability regions, rather than relying on visual inspection of individual FER curves.

We define the stability optimal initial LLR for each decoder as the minimizer of the AO $J(L_0)$,
\begin{equation}
L_{0,\star}^{\mathrm{BP2}} =
\arg\min_{L_0} \J^{\mathrm{BP2}}(L_0),
\qquad
L_{0,\star}^{\mathrm{BP4}}=
\arg\min_{L_0} \J^{\mathrm{BP4}}(L_0).
\label{eq:minimizer_def}
\end{equation}
The corresponding depolarizing parameters are obtained via the inverse likelihood mappings given in~\eqref{eq:bp2_e0} and~\eqref{eq:bp4_e0}, i.e.,
\[
\eps_{0,\star}^{\mathrm{BP2}}
=
\eps_{0}\!\left(L_{0,\star}^{\mathrm{BP2}}\right),
\qquad
\eps_{0,\star}^{\mathrm{BP4}}
=
\eps_{0}\!\left(L_{0,\star}^{\mathrm{BP4}}\right).
\]
%(cf.~Fig.~\ref{fig:bp4_several_eps_analysis}).
%%%%%%%%%%%%%%%%%%%%%%%%%%%%%%%%%%%
\subsection{Split Objective Decomposition}
To understand which noise regime dominates the AO, we partition the discretized operation region $\mathcal{G}=\{\eps_1,\ldots,\eps_T\}$ using a threshold $\eps_{\mathrm{split}}$. Applying the same definition as in~\eqref{eq:PAO_log} to each subset, we obtain:
\begin{equation}
\begin{aligned}
\J_{\mathrm{low}}(L_0)
&=
\sum_{\eps_t\le \eps_{\mathrm{split}}}
w_t\,\phi\!\left(\FER(\eps_t;L_0)\right), \\
\J_{\mathrm{high}}(L_0)
&=
\sum_{\eps_t> \eps_{\mathrm{split}}}
w_t\,\phi\!\left(\FER(\eps_t;L_0)\right).
\end{aligned}
\label{eq:PAO_split}
\end{equation}
where the original weights $w_t$ are preserved.
Since~\eqref{eq:PAO_log} defines $\J(L_0)$ as the uniform discrete expectation
over the selected grid $\mathcal{G}$, and the subsets
$\mathcal{G}_{\mathrm{low}}$ and $\mathcal{G}_{\mathrm{high}}$ form a disjoint
partition of $\mathcal{G}$, the total objective satisfies the (convex)
decomposition given by:
\begin{equation}
\J(L_0)=
\frac{\mathcal{G}_{\mathrm{low}}}{\mathcal{G}}\,\J_{\mathrm{low}}(L_0)+
\frac{\mathcal{G}_{\mathrm{high}}}{\mathcal{G}}\,\J_{\mathrm{high}}(L_0),
\label{eq:AO_convex_split}
\end{equation}
where $\mathcal{G}_{\mathrm{low}}$ and $\mathcal{G}_{\mathrm{low}}$ correspond to the cardinality points used when $\eps < \eps_{\mathrm{split}}$ and $\eps > \eps_{\mathrm{split}}$, resp.
%%%%%%%%%%%%%%%%%%%%%%%%%%%%%%
\section{Numerical Results}
\label{sec:results}
%%%%%%%%%%%%%%%%%%%%%%%%%%%%%%

We report BP2 and BP4 decoding results on the fixed OS GB$(126,28,126)$ 
construction of \cite{smiao_qbp_overcomplete2025} under i.i.d.\ depolarizing noise
with physical error rate $\eps$.
All experiments use identical Tanner graphs, stopping rules, number of iterations
$\ell^{\max}\in\{4,8\}$, and common random error realizations across initial LLRs 
values, ensuring fair comparisons.

%%%%%%%%%%%%%%%%%%%%%%%%%%%%%%%%%%%%%%%
\subsection{FER Performance with LLR Mismatch}
%%%%%%%%%%%%%%%%%%%%%%%%%%%%%%%%%%%%%%%

%%%%%%%%%%%%%%%%%%%%%%%%%%%%%

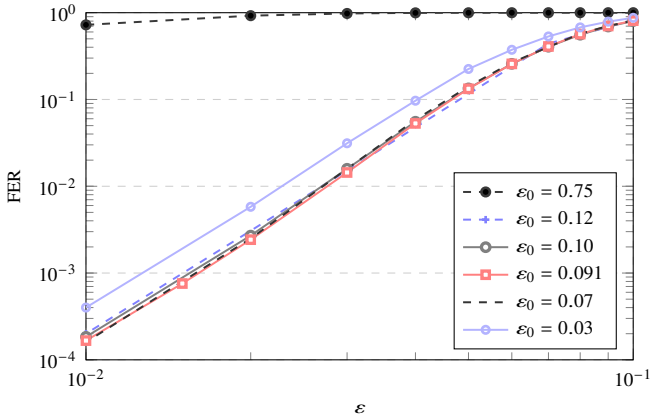
\begin{figure}[H]
\centering
\begin{tikzpicture}
\definecolor{set1}{RGB}{161,33,33}
\definecolor{set2}{RGB}{69,99,168}
\definecolor{set3}{RGB}{27,158,119}
\definecolor{set4}{RGB}{152,78,163}
\definecolor{set5}{RGB}{255,127,0}
\definecolor{set6}{RGB}{166,118,29}
\definecolor{set7}{RGB}{230,171,2}
\definecolor{set8}{RGB}{231,41,138}
\def\lw{0.75pt}
\begin{axis}[
    width=\columnwidth,
    height=0.70\columnwidth,
     yticklabel style={
    /pgf/number format/fixed,
    /pgf/number format/precision=3
    },
    every axis/.append style={font=\footnotesize},
    xmode=log, ymode=log,
    xmin=1e-2, xmax=1e-1,
    ymin=1e-4, ymax=1,
    ylabel near ticks,
    xlabel near ticks,
    xlabel={\scriptsize $\eps$},
    ylabel={\scriptsize FER},
    tick label style={font=\scriptsize},
    label style={font=\scriptsize},
    grid=major,
        grid style={dashed,lightgray!75},
   % major grid style={gray!25},
   % minor grid style={gray!12},
    legend style={
        font=\scriptsize,
        draw,
        fill=white,
        at={(0.97,0.03)},
        anchor=south east
    },
    legend cell align=left,
]
%%%%%%%%%%%%%%%%%%%%%%%%%%%%%%%%%%%%%%%%%
\addplot [color=black!10!darkgray, dashed, line width=\lw, mark=*, mark options={solid,scale=0.8, fill=black, line width=1pt}] file[]{figures/tikz_export/Wide_range_Prior_sweep_BP4/curve_01.dat};
\addlegendentry{$\eps_{0} = 0.75$}
%%%%%
\addplot [color=blue!50!white,dashed, line width=\lw, mark=+, mark options={solid,scale=0.7, fill=white, line width=1pt}] file[]
{figures/tikz_export/Wide_range_Prior_sweep_BP4/curve_02.dat};
\addlegendentry{$\eps_{0} = 0.12$}

\addplot [color=black!50!white, solid, line width=\lw, mark=*, mark options={solid,scale=0.8, fill=white, line width=1.25pt}] file[]
{figures/tikz_export/Wide_range_Prior_sweep_BP4/curve_04.dat};
\addlegendentry{$\eps_{0} = 0.10$}

\addplot [color=red!50!white, solid, line width=\lw, mark=square*, mark options={solid,scale=0.7, fill=white, line width=1.25pt}] file[]
{figures/tikz_export/Wide_range_Prior_sweep_BP4/curve_03.dat};
\addlegendentry{$\eps_{0} = 0.091$}
%%%%%%
\addplot [color=black!80!white,dashed, line width=\lw, mark=none, mark options={solid,scale=0.8, fill=white, line width=1pt}] file[]
{figures/tikz_export/Wide_range_Prior_sweep_BP4/curve_05.dat};
\addlegendentry{$\eps_{0} = 0.07$}
% black!20!white
\addplot [color=blue!30!white,solid, line width=\lw, mark=o, mark options={solid,scale=0.7, fill=white, line width=1pt}] file[]
{figures/tikz_export/Wide_range_Prior_sweep_BP4/GB_n126_k28_m126_BP4_iter4_mismatch_ep0_0p03.txt};
\addlegendentry{$\eps_{0} = 0.03$}

\end{axis}
\end{tikzpicture}
\vspace{-.5cm}
\caption{\footnotesize FER vs $\eps$ for BP4 with $\ell^{\max}=4$, obtaining similar performance for different $\eps_{0}$. Similar results (not shown) were observed for BP2.}
\label{fig:bp4_several_eps_analysis}
\end{figure}

Fig.~\ref{fig:bp4_several_eps_analysis} shows FER as a function of the channel depolarizing probability $\eps$ for different levels of LLR mismatch for BP4 and $\ell^{\max}=4$ iterations. We consider a wide range of $\eps_{0}$, from $\eps_{0}=0.03$ to $\eps_{0}=0.75$. The extreme case $\eps_{0}=0.75$ corresponds to $L_0=0$, i.e. no information about which error may have occurred. The other extreme value ($\eps_{0}=0.03$) is selected to span a broad initialization regime over which performance variations are comparable, at Monte Carlos uncertainty, thereby illustrating the width of the near optimal stability region. % Similar behavior is observed for BP2 (not shown for brevity).

%%%%%%%%%%%%%%%%%%%%%%%%%%%%%

The results in Fig.~\ref{fig:bp4_several_eps_analysis} show that the range $\eps_{0} \in \{0.07,0.12\}$ lead to near equivalent FER performance.

Throughout this paper, we use $\eps_{\mathrm{split}}=0.05$, as an arbitrary threshold (we will motivate this choice in the next section). The AO concept can be extended to more noise regime partitions and/or different thresholds (possibly combined with non uniform weights).

%%%%%%%%%%%%%%%%%%%%%%%%%%%%%%
\subsection{AO for BP4 and BP2}
%%%%%%%%%%%%%%%%%%%%%%%%%%%%%
%%%%%
We select the noise set to be $\mathcal{G}=\{0.15,\,0.13,\,0.11,\,0.10,\,0.09,\,0.07,\,0.05,\,0.03\}$ that intentionally provides more observations in the high noise regime (i.e., $0.11,0.13,0.15,\cdots$) relative to the chosen $\eps_{\mathrm{split}}$. Therefore, we choose $\eps_{\mathrm{split}}$ such that we allocate $25\%$ to the set $\mathcal{G}_{\mathrm{low}}$ and the remaining to $\mathcal{G}_{\mathrm{high}}$. During initial experiments (not shown here for brevity), when using the noise set used in Fig.~\ref{fig:bp4_several_eps_analysis}, i.e. $\eps \in \{0.01-0.1\}$, a dominance in $J(L_0)$ appeared, driven by the low noise regime. Therefore, we build this scenario to validate the hypothesis that the dominant source within this total set $\mathcal{G}$ would still be mostly influenced by the low noise regime.

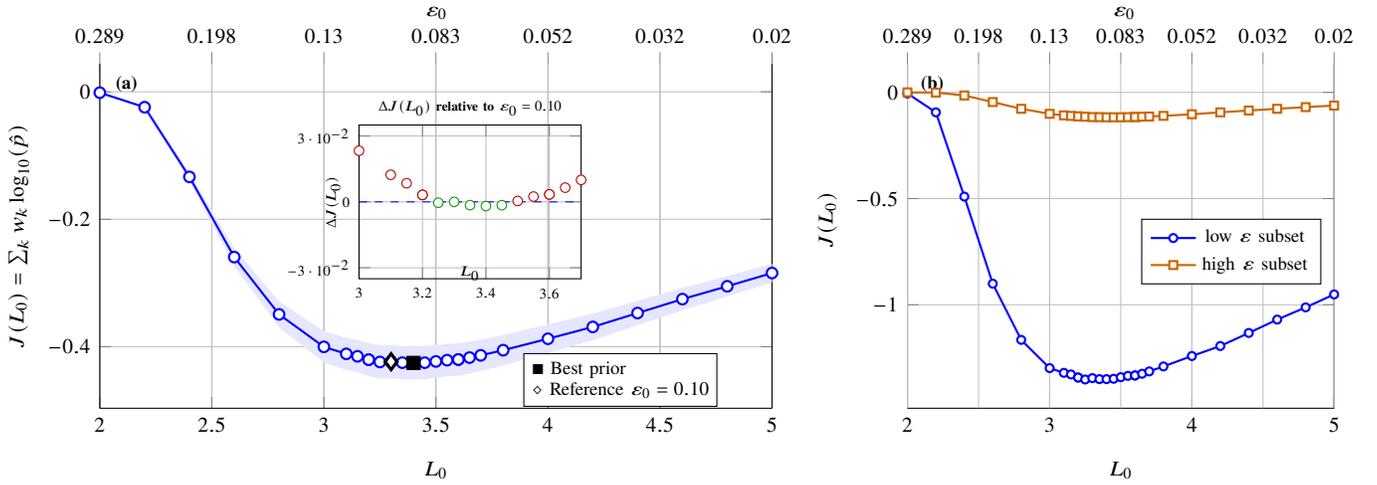
\begin{figure*}[t]
\centering
\begin{tikzpicture}

% ======================================================
% (a) aggregated objective
% ======================================================
\begin{axis}[
    name=mainA,
    width=0.58\textwidth,
    height=0.34\textwidth,
    at={(0,0)},
    anchor=south west,
    xlabel={\footnotesize $L_0$},
    axis x line*=bottom,
    ylabel={\footnotesize $J(L_0)=\sum_k w_k \log_{10}(\hat{p})$},
    xmin=2, xmax=5,
    grid=major,
    ticklabel style={font=\footnotesize},
    legend style={
        font=\scriptsize,
        at={(0.97,0.99)},
        anchor=north east,
        draw,
        fill=white,
    },
    extra x ticks={2,2.5,3,3.5,4,4.5,5},
    extra x tick labels={
    \epsfromLzeroBPfour{2},
    \epsfromLzeroBPfour{2.5},
    \epsfromLzeroBPfour{3},
    \epsfromLzeroBPfour{3.5},
    \epsfromLzeroBPfour{4},
    \epsfromLzeroBPfour{4.5},
    \epsfromLzeroBPfour{5}
    },
    extra x tick style={
    tick align=outside,
    tick pos=top,
    ticklabel pos=top,
    font=\tiny
    },
 %   extra x ticks={2,2.5,3,3.5,4,4.5,5},
%extra x tick labels={
%(\epsfromLzeroBPfour{2}),
%(\epsfromLzeroBPfour{2.5}),
%(\epsfromLzeroBPfour{3}),
%(\epsfromLzeroBPfour{3.5}),
%(\epsfromLzeroBPfour{4}),
%(\epsfromLzeroBPfour{4.5}),
%(\epsfromLzeroBPfour{5})
%},
%    extra x tick style={tick label style={yshift=-10pt,font=\scriptsize}},
]
% CI band 
\addplot[name path=cihigh, draw=none]
  table {figures/tikz_export/bp4_ep0_study/data_txt/BP4_J_vs_L0_CI_high.txt};

\addplot[name path=cilow, draw=none]
  table {figures/tikz_export/bp4_ep0_study/data_txt/BP4_J_vs_L0_CI_low.txt};

\addplot[fill=blue!10, draw=none]
  fill between[of=cihigh and cilow];

%Objective mean 
\addplot[blue, thick, mark=*,
    mark options={solid,fill=white},, mark size=2pt]
  table {figures/tikz_export/bp4_ep0_study/data_txt/BP4_J_vs_L0_mean.txt};
%\addlegendentry{Aggregated Objective}

%  Best 
\addplot[only marks, mark=square*, mark size=2.5pt, black]
  table {figures/tikz_export/bp4_ep0_study/data_txt/BP4_J_vs_L0_best.txt};

% Reference
\addplot[only marks, mark=diamond*, mark size=3pt, black, fill=white, very thick]
  table {figures/tikz_export/bp4_ep0_study/data_txt/BP4_J_vs_L0_ref.txt};

% Annotation instead of legend entries that were not working....
\node[
    anchor=north west,
    font=\scriptsize,
    draw,
    fill=white,
    align=left,
    inner sep=3pt,
] at (rel axis cs:0.63,0.16) {
$\blacksquare$ Best prior\\
$\diamond$ Reference $\eps_0=0.10$
};
\node[anchor=north west,font=\bfseries \scriptsize] at (rel axis cs:0.01,0.99) {(a)};

\end{axis}
\node[font=\footnotesize]
    at (rel axis cs:0.5,1.23) {$\eps_{0}$};
%%%%%%%%%%%%%%%%%%%%%%%%%%
% INSET: ΔJ(L0) 
% ======================================================
\begin{axis}[
    width=0.25\textwidth,
    height=0.20\textwidth,
    at={(mainA.south west)},
    anchor=south west,
    xshift=0.19\textwidth,
    yshift=0.095\textwidth,
    xmin=3.0, xmax=3.7,
    ymin=-0.035, ymax=0.035,
    ytick={-0.03,0,0.03},
    grid=major,
    xlabel={$L_0$},
    xlabel style={
    at={(axis description cs:0.5,0.14)},
    anchor=north,
    font=\tiny
    },
    ylabel={$\Delta J(L_0)$},
    ylabel style={
    at={(axis description cs:-0.04,0.5)},
    anchor=south,
    rotate=0,
    font=\tiny
    },
    title={\bfseries \tiny $\Delta J(L_0)$ relative to $\eps_0=0.10$},
    title style={yshift=-6pt},
    ticklabel style={font=\tiny},
    label style={font=\tiny},
    title style={font=\scriptsize},
    xtick={3.0,3.2,3.4,3.6},
    scaled ticks=false,
    enlargelimits=false,
]

% ΔJ < 0 (better than reference)
\addplot[
    green!60!black,
    only marks,
    mark=*,
    mark options={solid,fill=white},
    mark size=1.8pt
]
table {figures/tikz_export/bp4_ep0_study/data_txt/BP4_dJ_vs_L0_neg.txt};

% ΔJ > 0 (worse than reference)
\addplot[
    red!70!black,
    only marks,
    mark=*,
    mark options={solid,fill=white},
    mark size=1.8pt
]
table {figures/tikz_export/bp4_ep0_study/data_txt/BP4_dJ_vs_L0_pos.txt};

% zero reference
%\addplot[dashed, blue]
%coordinates {(3.0,0) (3.7,0)};
\addplot[dashed, blue] {0};
\end{axis}
%%%%%%%%%%%%%%%%%%%%%%%%%
% ======================================================
% (b) split into elow and e high
% ======================================================
\begin{axis}[
    width=0.4\textwidth,
    height=0.34\textwidth,
    at={(mainA.east)},
    anchor=west,
    xshift=18mm,
    ticklabel style={font=\footnotesize},
    xlabel={\footnotesize $L_0$},
    axis x line*=bottom,
    ylabel={\footnotesize $J(L_0)$},
    ylabel style={xshift=6pt},
    xmin=2, xmax=5,
    grid=major,
    legend style={font=\scriptsize, at={(0.97,0.55)}, anchor=north east},
    extra x ticks={2,2.5,3,3.5,4,4.5,5},
extra x tick labels={
\epsfromLzeroBPfour{2},
\epsfromLzeroBPfour{2.5},
\epsfromLzeroBPfour{3},
\epsfromLzeroBPfour{3.5},
\epsfromLzeroBPfour{4},
\epsfromLzeroBPfour{4.5},
\epsfromLzeroBPfour{5}
},
    extra x tick style={
    tick align=outside,
    tick pos=top,
    ticklabel pos=top,
    font=\tiny
    },
]

\addplot[blue, thick, mark=*,
    mark options={solid,fill=white},, mark size=1.5pt]
  table {figures/tikz_export/bp4_ep0_study/data_txt/BP4_J_vs_L0_low_eps.txt};
\addlegendentry{low $\eps$ subset}

\addplot[orange!80!black, thick, mark=square*,
    mark options={solid,fill=white}, mark size=1.5pt]
  table {figures/tikz_export/bp4_ep0_study/data_txt/BP4_J_vs_L0_high_eps.txt};
\addlegendentry{high $\eps$ subset}

\node[anchor=north west,font=\bfseries \scriptsize] at (rel axis cs:0.01,0.99) {(b)};

\end{axis}
\node[font=\footnotesize]
    at (rel axis cs:2.4,1.23) {$\eps_{0}$};
\end{tikzpicture}

\caption{\footnotesize
BP4 AO vs mismatched LLR. The blue shaded region denotes the AO standard deviation. 
(a) The optimum corresponds to a region of near equivalent priors. The inset shows that $\eps_{0}=0.10$ is a pragmatic choice for BP4 over the simulated noise regime;
(b) Sensitivity is dominated by the low noise subset.
}
\label{fig:bp4_mismatch_tuning_split_study}
\end{figure*}

\begin{figure*}[t]
\centering
\begin{tikzpicture}

% ======================================================
% (a) aggregated objective
% ======================================================
\begin{axis}[
    name=mainA,
    width=0.58\textwidth,
    height=0.34\textwidth,
    at={(0,0)},
    anchor=south west,
    xlabel={\footnotesize $L_0$},
    axis x line*=bottom,
    ylabel={\footnotesize $J(L_0)=\sum_k w_k \log_{10}(\hat{p})$},
    xmin=2, xmax=5,
    ticklabel style={font=\footnotesize},
    grid=major,
    legend style={
        font=\scriptsize,
        at={(0.97,0.99)},
        anchor=north east,
        draw,
        fill=white,
    },
    extra x ticks={2,2.5,3,3.5,4,4.5,5},
extra x tick labels={
\epsfromLzeroBPtwo{2}),
\epsfromLzeroBPtwo{2.5},
\epsfromLzeroBPtwo{3},
\epsfromLzeroBPtwo{3.5},
\epsfromLzeroBPtwo{4},
\epsfromLzeroBPtwo{4.5},
\epsfromLzeroBPtwo{5}
},
    extra x tick style={
    tick align=outside,
    tick pos=top,
    ticklabel pos=top,
    font=\tiny
    },
]
% CI band 
\addplot[name path=cihigh, draw=none]
  table {figures/tikz_export/bp2_ep0_study/data_txt/BP2_J_vs_L0_CI_high.txt};

\addplot[name path=cilow, draw=none]
  table {figures/tikz_export/bp2_ep0_study/data_txt/BP2_J_vs_L0_CI_low.txt};

\addplot[fill=blue!10, draw=none]
  fill between[of=cihigh and cilow];

%Objective mean 
\addplot[blue, thick, mark=*,
    mark options={solid,fill=white}, mark size=2pt]
  table {figures/tikz_export/bp2_ep0_study/data_txt/BP2_J_vs_L0_mean.txt};
%\addlegendentry{Aggregated Objective}

%  Best 
\addplot[only marks, mark=square*, mark size=2.5pt, black]
  table {figures/tikz_export/bp2_ep0_study/data_txt/BP2_J_vs_L0_best.txt};

% Reference
\addplot[only marks, mark=diamond*, mark size=3pt, black, fill=white, very thick]
  table {figures/tikz_export/bp2_ep0_study/data_txt/BP2_J_vs_L0_ref.txt};

% Annotation instead of legend entries that were not working....
\node[
    anchor=north west,
    font=\scriptsize,
    draw,
    fill=white,
    align=left,
    inner sep=3pt
] at (rel axis cs:0.63,0.16) {
$\blacksquare$ Best prior\\
$\diamond$ Reference $\eps_0=0.10$
};
\node[anchor=north west,font=\bfseries \scriptsize] at (rel axis cs:0.01,0.99) {(a)};
\end{axis}
%%%%%%%%%%%%%%%%%%%%%%%%%%
\node[font=\footnotesize]
    at (rel axis cs:0.5,1.23) {$\eps_{0}$};
% INSET: ΔJ(L0) 
% ======================================================
\begin{axis}[
    width=0.25\textwidth,
    height=0.20\textwidth,
    at={(mainA.south west)},
    anchor=south west,
    xshift=0.12\textwidth,
    yshift=0.11\textwidth,
    xmin=2.6, xmax=3.1,
    ymin=-0.035, ymax=0.035,
    ytick={-0.03,0,0.03},
    grid=major,
    xlabel={$L_0$},
    xlabel style={
    at={(axis description cs:0.5,0.14)},
    anchor=north,
    font=\tiny
    },
    ylabel={$\Delta J(L_0)$},
    ylabel style={
    at={(axis description cs:-0.04,0.5)},
    anchor=south,
    rotate=0,
    font=\tiny
    },
    title={\bfseries \tiny $\Delta J(L_0)$ relative to $\eps_0=0.10$},
    title style={yshift=-6pt},
    ticklabel style={font=\tiny},
    label style={font=\tiny},
    title style={font=\scriptsize},
    scaled ticks=false,
    enlargelimits=false,
]

% ΔJ < 0 (better than reference)
\addplot[
    green!60!black,
    only marks,
    mark=*,
    mark options={solid,fill=white},
    mark size=1.8pt
]
table {figures/tikz_export/bp2_ep0_study/data_txt/BP2_dJ_vs_L0_neg.txt};

% ΔJ > 0 (worse than reference)
\addplot[
    red!70!black,
    only marks,
    mark=*,
    mark options={solid,fill=white},
    mark size=1.8pt
]
table {figures/tikz_export/bp2_ep0_study/data_txt/BP2_dJ_vs_L0_pos.txt};

% zero reference
%\addplot[dashed, blue]
%coordinates {(2.6,0) (3.2,0)};
\addplot[dashed, blue] {0};
\end{axis}
%%%%%%%%%%%%%%%%%%%%%%%%%
% ======================================================
% (b) split into elow and e high
% ======================================================
\begin{axis}[
    width=0.4\textwidth,
    height=0.34\textwidth,
    at={(mainA.east)},
    anchor=west,
    xshift=18mm,
    xlabel={\footnotesize $L_0$},
    axis x line*=bottom,
    ylabel={\footnotesize $J(L_0)$},
    ylabel style={xshift=6pt},
    xmin=2, xmax=5,
    ticklabel style={font=\footnotesize},
    grid=major,
    legend style={font=\scriptsize, at={(0.97,0.55)}, anchor=north east},
    extra x ticks={2,2.5,3,3.5,4,4.5,5},
extra x tick labels={
\epsfromLzeroBPtwo{2},
\epsfromLzeroBPtwo{2.5},
\epsfromLzeroBPtwo{3},
\epsfromLzeroBPtwo{3.5},
\epsfromLzeroBPtwo{4},
\epsfromLzeroBPtwo{4.5},
\epsfromLzeroBPtwo{5}
},
    extra x tick style={
    tick align=outside,
    tick pos=top,
    ticklabel pos=top,
    font=\tiny
    },
]

\addplot[blue, thick, mark=*,
    mark options={solid,fill=white}, mark size=1.5pt]
  table {figures/tikz_export/bp2_ep0_study/data_txt/BP2_J_vs_L0_low_eps.txt};
\addlegendentry{low $\eps$ subset}

\addplot[orange!80!black, thick, mark=square*,
    mark options={solid,fill=white}, mark size=1.5pt]
  table {figures/tikz_export/bp2_ep0_study/data_txt/BP2_J_vs_L0_high_eps.txt};
\addlegendentry{high $\eps$ subset}

\node[anchor=north west,font=\bfseries \scriptsize] at (rel axis cs:0.01,0.99) {(b)};

\end{axis}
\node[font=\footnotesize]
    at (rel axis cs:2.4,1.23) {$\eps_{0}$};
\end{tikzpicture}

\caption{\footnotesize
BP2 AO vs mismatched LLR. The blue shaded region denotes the AO standard deviation. 
(a) The optimum corresponds to a region of near equivalent priors. The inset shows that $\eps_{0}=0.10$ is still in the limit to be considered within the optimum zone.
(b) Similarly to the case for BP4, sensitivity is also dominated by the low noise subset.
}
\label{fig:bp2_mismatch_tuning_split_study}
\end{figure*}
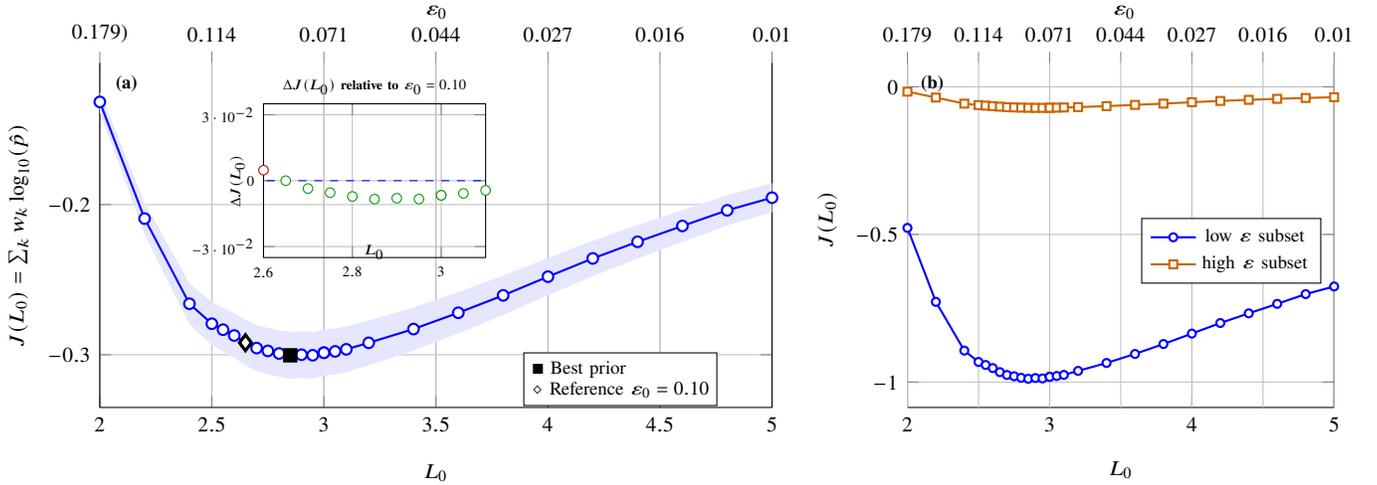

We found that, over the operating set defined by $\mathcal{G}$, the optimal initial LLR occurs at:
\begin{equation*}
\begin{aligned}
\left(
\eps_{0,\star}^{\mathrm{BP4}},\;
L_{0,\star}^{\mathrm{BP4}}
\right)
&\approx
\left(
0.091,\; 3.4
\right), \\[4pt]
\left(
\eps_{0,\star}^{\mathrm{BP2}},\;
L_{0,\star}^{\mathrm{BP2}}
\right)
&\approx
\left(
0.082,\; 2.85
\right).
\end{aligned}
\label{eq:optimal_pairs}
\end{equation*}
%%%%%%%%%%%%%%
Figures~\ref{fig:bp4_mismatch_tuning_split_study} and \ref{fig:bp2_mismatch_tuning_split_study} show the AO $\J(L_0)$ in \eqref{eq:PAO_log} for BP4 and BP2, respectively. The AO exhibits a near flat region for $L_0\in[3.2,3.5]$ 
in the case of BP4, whereas for BP2, the corresponding region spans $L_0\in[2.6,3.2]$.
Within these intervals, variations in $\J(L_0)$ are comparable to Monte Carlo uncertainty, indicating that finite iteration performance is insensitive to minor tuning of $\eps_{0}$ around the optimum value.
The choice of the initial LLR influences how rapidly messages approach saturation in early iterations. On OS graphs with dense short cycles, correlated feedback may accumulate before the decoder has effectively incorporated information from the full graph. Initial LLR mismatch alters this early behavior without changing the decoding objective itself, leading to computable differences in finite iteration FER. This interpretation reinforces the argument that on redundant loopy OS graphs, the assumed initial LLR acts as a practical regularization parameter under the chosen number of iterations $\ell$.

 %%%%%%%%%%%%%%%%%%%%%%%%%%%%%%%%%%%%%%%%%%%%%%
%%%%%%%%%%%%%%%%%%%%%%%%%%%%%%%%%%%%%%%%%%%%%%

%%%
The insets in Figs.~\ref{fig:bp4_mismatch_tuning_split_study}(a) and
\ref{fig:bp2_mismatch_tuning_split_study}(a) plot
$\Delta\J(L_0)=\J(L_0)-\J(L_{0,\mathrm{ref}})$
where $J(L_{0,\mathrm{ref}})$ corresponds to the reference $\eps_{0}=0.10$.
We observe that, close to the optimal region, differences are comparable
to statistical uncertainty. From a practical standpoint, this supports selecting a representative value within the stability region
(e.g., $\eps_{0}=0.10$ as reported in\cite{smiao_qbp_overcomplete2025} for BP4) rather than attempting fine grained optimization or matching it to the channel ($\eps_0=\eps)$.

%%%%%
The split objective decomposition in Figs.~\ref{fig:bp4_mismatch_tuning_split_study}(b) and ~\ref{fig:bp2_mismatch_tuning_split_study}(b) validates our initial hypothesis, confirming that the main component of the initial sensitivity of the LLR value comes from the low noise regime, where decoding failures are rare and therefore highly sensitive to early iteration dynamics.
In contrast, the high noise contribution is less significant
with respect to $\eps_{0}$ and although $\mathcal{G}_{\mathrm{low}} < \mathcal{G}_{\mathrm{high}}$, 
the ow noise subset still dominates the variation in $\J(L_0)$.

%%%%
A final observation is noticed on the fact that the sensitivity to the initial LLR weakens as the number of iterations $\ell$ increases. This confirms that initial LLR mismatch modifies early message dynamics on loopy OS graphs but does not alter the outcome of the decoding process.
%%%%%%%%%%%%%%%%%%%%%%%%%%%

\section{Conclusions}
%%%%%%%%%%%%

In this paper, we investigated the effect of LLR initialization on the performance of BP2 and BP4 decoding of QLDPC codes with overcomplete check matrices. We focused on relatively low number of iterations and showed that moderate mismatch in initialization LLR can significantly improve FER performance. The observed sensitivity appears for both BP2 and BP4, indicating that it reflects a general finite iteration phenomenon of BP on redundant (overcomplete) check matrices and corresponding loopy Tanner graphs. We then showed that the performance is relatively insensitive to small changes in mismatch around the optimum. 

In this paper, we also proposed an aggregated objective function to quantify the effect of LLR mismatch. Our results  show two important aspects of LLR mismatch for decoding QLDPC codes. Firstly, the proposed function can help predict the optimum initialization region, thereby relaxing the requirement of extensive numerical simulations. Secondly, the results confirm the interpretation of LLR initialization as a finite iteration regularization parameter, controlling the scale of early BP message magnitudes and feedback reinforcement along short cycles, instead of as a quantity that must precisely match the actual channel.

\balance

%%%%%%%%%%%%%%%%%%%%%%%%%%%%%%%%%%
%\newpage
\bibliographystyle{IEEEtran}
\bibliography{references}
\end{document}